\documentclass[12pt,epsfig,preprint]{aastex}

\shortauthors{B. Paul et al.}
\shorttitle{EXO 0531-66}

\begin{document}

\title{
Timing and spectral studies of the transient X-ray
pulsar EXO~053109-6609.2 with ASCA and Beppo-SAX}

\author{B. Paul\altaffilmark{1}, S. N. A. Jaaffery\altaffilmark{2},
S. Naik\altaffilmark{1,3} and P. C. Agrawal\altaffilmark{1}}

\altaffiltext{1}{Department of Astronomy and Astrophysics,
Tata Institute of Fundamental Research, Homi Bhabha road, Mumbai,
400\,005, India. bpaul@tifr.res.in}

\altaffiltext{2}{Department of Physics, College of Science,
M. L. Sukhadia University, Udaipur,  Rajasthan, 313\,001, India.}

\altaffiltext{3}
{Department of Physics University College Cork, Cork, Ireland}

\date{}

\begin{abstract}

We report timing and spectral properties of the transient
Be X-ray pulsar EXO~053109--6609.2 studied using observations
made with the ASCA and Beppo-SAX observatories.
Though there must have been at least one spin-down episode
of the pulsar since its discovery, the new pulse period
measurements show a monotonic spin-up trend since 1996.
The pulse profile is found to have marginal energy dependence.
There is also evidence for strong luminosity dependence of the pulse
profile, a single peaked profile at low luminosity that
changes to a double peaked profile at high luminosity. This
suggests a change in the accretion pattern at certain luminosity
level. The X-ray spectrum is found to consist of a simple
power-law with photon index in the range of 0.4--0.8. At high
intensity level the spectrum also shows presence of weak iron
emission line.

\end{abstract}


\section{Introduction}

The two Magellanic Clouds have larger number density of High
Mass X-ray Binaries (HMXB) compared to our galaxy. The Small
Magellanic Cloud (SMC) with a mass of about 1\% of our
Galaxy has a total of about 70 HMXBs and HMXB candidates,
almost comparable to that in our galaxy (Haberl \& Sasaki 2000;
Yokogawa et al. 2003). Though the HMXB number density in the
Large Magellanic Cloud (LMC) is not as high as that in SMC,
it is still significantly higher than the Galactic
value (Sasaki, Haberl \& Pietsch 2000). In-spite of a large number
of HMXB pulsar objects in the two Magellanic Clouds,
individual objects have not been studied in great detail
except for SMC~X-1 and LMC~X-4. There are
four HMXB pulsars known in the LMC, of which LMC~X-4 is the best
studied. Several of these sources show a very large X-ray luminosity
in their high states, close to or exceeding the Eddington limit for
a 1 M$_{\sun}$ object. Sources, in which a strong local absorption
column is absent, nature of the soft spectral component can be studied
in detail. Due to a low galactic absorption column density towards the
LMC and SMC, these objects are also suitable for study of the local
absorbing material. A soft excess in the X-ray
spectrum, now detected in several accreting pulsars can be a
common feature of the HMXB pulsars. But since most of the HMXB
pulsars are in the Galactic plane, large line of sight absorption
usually makes it difficult to detect it. It is therefore, important
to study details of the individual HMXBs in the LMC and SMC.

The transient Be X-ray binary source EXO~053109--6609.2 was discovered
with the EXOSAT observatory in 1983 (Pietsch, Rosso \& Dennerl 1989).
Hard X-ray emission from this source was detected with the SL2 XRT
experiment (Hanson et al. 1989) and its intensity variations were
studied in detail with the ROSAT (Haberl, Dennerl, \& Pietsch 1995).
X-ray pulsations were also discovered with the ROSAT (Dennerl,
Haberl \& Pietsch 1995). This pulsar is located at a distance
of 17$\arcmin$ from the luminous binary X-ray pulsar LMC~X-4,
and has a pulse period of 13.7 s, which is also close to the 13.5 s
pulse period of LMC~X-4. During a Beppo-SAX observation of LMC~X-4,
EXO~053109--6609.2 was in the field of view of the narrow field
instruments of Beppo-SAX, and pulsations were detected along with
a positive local period derivative (Burderi et al. 1998). The source
was also observed with the XMM in 2000 during which it was in a relatively
bright state (Haberl, Dennerl, \& Pietsch 2003).
The optical counterpart
is one of the two close double Be-stars (Stevens, Coe and Buckley 1999)

In the subsequent sections we present the archival observations
that we have used to study this source (\S 2), results from
temporal and spectral analysis of the data (\S 3) and follow
up this with a discussion on the nature of this pulsar (\S 4).

\section{Observations}

We have analysed archival data from four observations of the pulsar
EXO~053109--6609.2. Three of these observations were made with the
Beppo-SAX observatory and one with the ASCA.

The field of EXO~053109--6609.2 was covered by ASCA-GIS during several
observations of nearby sources. In one observation of LMC~X-4, made
during 1996 May 24--26, we found EXO~053109--6609.2 to be bright enough
for temporal and spectral investigation. It was also present in the
field of view of the Beppo-SAX narrow field instruments during two
observations of LMC~X-4 in 1997 (March 13--15) and 1998 (October 20--22)
and in another one observation of the LMC in 2000 (January 01--02).
EXO~053109--6609.2 was in a transient outburst during the 1997
observation and the pulsation properties during this period have been
reported earlier (Burderi et. al. 1998).
We have found 36 ks of useful exposure
with ASCA-GIS, 117, 82 and 32 ks with SAX-MECS (1997, 1998 and 2000
respectively), and 42, 31 and 15 ks with SAX-LECS.
For more details of these instruments,
refer to Tanaka, Inoue \& Holt (1994, ASCA) and Boella et al. (1997, SAX).

\section{Analysis and results}

We have used the standard data selection criteria of both the
observatories. The count rates and spectra of the source were obtained
from circular regions of radius 6$\arcmin$ for ASCA-GIS,
3$\arcmin$ for SAX-MECS and 2$\arcmin$ for SAX-LECS.
The corresponding background spectra
were extracted from source free nearby regions in the respective FOVs.

\subsection{Timing analysis}

Light curves from these observations were extracted with time
resolution of 0.25 s for SAX LECS and MECS and 0.50 s for ASCA-GIS.
Barycenter corrections were applied
to the photon arrival times in the event files before extraction of
light curves. The pulse folding and $\chi^{2}$ maximisation search
near the expected pulse period of 13.7 s clearly showed presence of
pulsations in all the observations. The new pulse period measurements
with ASCA-GIS and Beppo-SAX LECS/MECS are given in Table 1 and
plotted in Figure 1.

Pulse profiles were created from background subtracted light curves
in different energy bands. The pulse profiles of all but the 1997
Beppo-SAX observation are shown in Figure 2 along with the dates of
each observation and the 2--10 keV luminosity of the source during the
corresponding observation. There is a hint of slight energy dependence
of the pulse profile in the ASCA observation. The pulse profile
clearly shows a luminosity dependence; a single peaked
profile at low luminosity (ASCA 1996, SAX 1998, top panels of
Figure 2.) and a double peaked profile at high luminosity.
The pulse fraction is in the range of 30--60\%.

The pulse period history of EXO~053109--6609.2, obtained by combining
the new observations and the earlier reported measurements with
the ROSAT, Beppo-SAX and XMM observatories are plotted in Figure 1.
An overall spin-down trend during the last five years and some
local variation from this
trend can be seen in the figure. The orbital elements of this binary
system is not known, and we note that an orbital motion of the neutron
star with a velocity of a few hundred km s$^{-1}$ can also result in the
observed changes in the local pulse period measurements.

\subsection{Spectral analysis}

\subsubsection{ASCA}

Suitable binning was applied to all the energy spectra before spectral
fitting. We first fitted the ASCA-GIS spectrum with a simple model
consisting of a power-law component and line of sight absorption
with solar abundance of metals.
The source was in a low intensity state during this ASCA observation.
We rebinned the spectrum to 12 spectral bins in the 0.7--9.0 keV
range. A photon index of 0.46 and equivalent hydrogen column density of
$1.1\times10^{21}$ atoms cm$^{-2}$ was obtained with a reduced
$\chi^{2}$ of 0.23 for
9 degrees of freedom. Assuming a distance of 55 kpc to the LMC, the
X-ray luminosity during the ASCA observation was $1.2\times10^{36}$
erg s$^{-1}$. No signature of iron fluorescence was detected and
a broad binning scheme needed due to the poor statistics also does
not allow us to put a strong upper limit on the equivalent width
of a line feature at 6.4 keV. The spectrum measured with ASCA-GIS
is shown in Figure 3 along with the best fit model spectrum and
the residuals.

\subsubsection{SAX}

In all the three Beppo-SAX observations, the source EXO~053109--6609.2
was at a distance of about 17$\arcmin$ away from the centre of the
field of view. Therefore, it was necessary to consider the smaller
collecting area of the mirrors at large off-axis angles. For spectral
analysis of MECS spectra at an off-axis angle of 17$\arcmin$, we have
taken an average of the effective areas at 14$\arcmin$ and 20$\arcmin$,
which are available for the MECS. This is likely to introduce a few
percent uncertainty in the measurements of spectral parameters over the
statistical errors. At low energies, the relative loss of effective
area for different off-axis angles is not energy dependent. Therefore,
the LECS spectra were fitted with the on-axis effective area curve,
and a relative normalisation with respect to the MECS was used.
All the three Beppo-SAX observations were fitted with the same spectral
model described above. In the 1997 and 2000 observations, there was
some signature of an iron emission line, and a separate narrow Gaussian
line at 6.4 keV was included in the model. The equivalent width of the
iron line during these two observations were 130 $\pm$ 50 eV and
180 $\pm$ 160 eV respectively. The spectral parameters along
with the reduced $\chi^{2}$ obtained are given in Table 2. The X-ray
luminosities in the 2.0--10.0 keV band during the three SAX observations
are $6.1\times10^{36}$, $3.0\times10^{35}$, and $1.5\times10^{36}$
erg s$^{-1}$ during the 1997, 1998 and 2000 observations respectively.
The spectra measured with Beppo-SAX LECS and MECS instruments are shown
in Figure 3.

\section{Discussion}

The pulsar EXO~053109--6609.2, like many other HMXB pulsars with Be
star companions, is believed to be a transient source. However, we
have found this pulsar to be bright, at a luminosity level of 
$3 \times 10^{35}$ erg s$^{-1}$ to $6 \times 10^{36}$ erg s$^{-1}$
in most of the X-ray observations made since 1993 with the low
background imaging instruments. Several ROSAT observations during
March to May 1993 also found the source to have significant X-ray intensity
(Haberl et al. 1995). In this respect, this pulsar is more akin
to 4U~1907+09 (In't Zand 1999, Mukerjee et al. 2001), which also
shows intensity variations of similar scale but without any
obvious transient phenomena. Its long term X-ray intensity behaviour
is different from that of the Be-Star transients like EXO~2030+375 (Parmar,
White \& Stella 1989).
The peak X-ray luminosity of $6 \times 10^{36}$ erg s$^{-1}$
observed in 1997 is however, much smaller than that of the
pulsars SMC~X-1 and LMC~X-4 (Paul et al. 2002), which have high
mass companion stars in very close orbits. This indicates a large
size orbit and hence orbital period. The archival X-ray observations
are not sufficiently long to search for pulse arrival time delay
that can help to determine the orbital parameters. Analysing the
RXTE-ASM light curve of EXO~053109--6609.2 we have found a peak
in the power spectrum at 183 days and its integer multiples.
Intensity variations of the bright nearby source LMC~X-4 is
unlikely to create this because the 30.5 day superorbital intensity
variation of LMC~X-4 is not reflected in the power spectrum of
EXO~053109--6609.2. However, since this is very close to a half-year,
yearly observational effects cannot be ruled out.

Combining the present observations with the earlier measurements
of pulse periods with the ROSAT and the Beppo-SAX, we have obtained a
period history of the pulsar, shown in Figure 1. It shows a clear
monotonic spin-up trend since the ASCA observation in 1996, though
not with a constant period derivative. However, certainly there were
spin-down episodes between the 1991 ROSAT observation and the 1996
ASCA observation. A spin-down was also observed during the 1997
Beppo-SAX observation itself, which is a surprise considering the
high intensity level during this observation. The last 5 years of
pulse period measurements presented here shows a spin-up of the
neutron star. This suggests that accretion onto the neutron star was in
progress for at least a significant part of this period. However,
we cannot rule out a significant binary motion related component
in the observed pulse period variations.

The X-ray emission from the polar cap regions can have different
geometric pattern at different mass accretion rates. In the
transient X-ray pulsar EXO 2030+375, a change in beaming pattern,
from a fan beam in high luminosity state to a pencil-beam in
low luminosity state was observed with the EXOSAT (Parmar et al.
1989). The pulse profiles of EXO~053109--6609.2 at different
luminosity levels presented here and elsewhere
(Burderi et al. 1998, Haberl et al. 2003)
clearly indicate a
dependence on the mass accretion rate. The pulse profile is single peaked
and nearly sinusoidal in low intensity state (Beppo-SAX : 1998), and it has
two distinct well separated peaks or an admixture of the above two
during other observations.
It is possible that in EXO~053109--6609.2 also,
the X-ray beam pattern changes from a pencil beam at low luminosity
to a fan beam at high luminosity. The change in beam pattern
probably takes place at a luminosity level of $\sim10^{36}$
erg s$^{-1}$.

During the ASCA observation in 1996 and the Beppo-SAX observation
in 1998, the source was in low intensity state and the
spectral parameters are not constrained well from these observations.
However, the spectral characteristics reported here from the
other three observations indicate intrinsic spectral variability in the
pulsar. The power-law component of the spectrum is softer at higher
luminosity.

The XMM spectrum of this source unambiguously showed the presence
of a soft spectral component in excess of a partially absorbed
hard power-law extended to lower energies (Haberl et al. 2003).
The same observation also showed that the pulse profile is different in the
soft and the hard X-ray bands. This indicates that the soft component
may have a different origin, or at least the geometry of emission is
different in different energy ranges. Presence of narrow features
in the soft excess prompted Haberl et al. (2003) to use a hot plasma
model that yields a temperature of 0.1 keV and emission measure of
4 $\times$ 10$^{61}$ cm$^{-3}$. Such a plasma is unlikely to show
pulsations due to its large physical size and large cooling time
scale, contrary to what is seen in the 0.15-0.4 keV band. A blackbody
type component with separate low energy emission lines as
seen in Her X-1 (Endo et al. 2000) may be a better description of
the soft excess of this pulsar. Several other accreting X-ray pulsars with
relatively low line of sight absorption show soft excess (Paul et al.
2002 and references therein). In some sources the soft excess probably
originates in a part of the X-ray irradiated inner accretion disk
(Endo et al. 2000; Naik \& Paul 2004). From independent analysis of
the XMM spectra we have found that in EXO~053109--6609.2, the total
luminosity of unabsorbed X-ray emission in the soft excess is
$1.1 \times 10^{36}$ erg s$^{-1}$, which is only about 1\% of the
total emission in the 1--10 keV energy band.
The ratio of the soft excess flux to the
total flux is smaller than that of SMC~X-1 (3.6\%), LMC~X-4 (6.4\%)
and Her~X-1 (10\%). In this pulsar, the size of the
black body emission region is required to be only about $10^{7}$ cm
and, therefore, can easily be accommodated in inner part of
the accretion disk heated by the hard X-rays.

\begin{acknowledgements}

We thank an anonymous referee for valuable suggestions.
This research has made use of data obtained from the High Energy
Astrophysics Science Archive Research Center (HEASARC), provided by
NASA's Goddard Space Flight Center. SNAJ thanks the ASTROSAT programme
through which a visit to TIFR was supported.

\end{acknowledgements}

\clearpage

\begin{deluxetable}{cccc}
\footnotesize
\tablenum{1}
\tablecaption{The pulse period measurements of EXO~053109--6609.2 \label{tbl-1}}
\tablewidth{0pt}
\tablehead{
\colhead{Date of Obs. (MJD)}&\colhead{Observatory - Instrument}&\colhead{Pulse period (s)}&\colhead{Reference}\\}
\startdata

48560.16 & ROSAT - PSPC &     13.67133 $\pm$ 0.00005 & Dennerl et al. (1995)\\
50228.12 & ASCA - GIS &       13.67875 $\pm$ 0.00012 & Present work\\
50520.00  & SAX - MECS &      13.67590 $\pm$ 0.00008 & Burderi et. al. (1998)\\
51107.64  & SAX - MECS &      13.670674 $\pm$ 0.000026 & Present work\\
51600.72  & SAX - MECS &      13.66884 $\pm$ 0.00004 & Present work\\
51824.56  & XMM - MOS+PN &    13.66817 $\pm$ 0.00001 & Haberl et al. (2003)\\
\enddata

\end{deluxetable}

\begin{deluxetable}{lccccc}
\footnotesize
\tablenum{2}
\tablecaption{Spectral parameters of EXO~053109--6609.2 during different observations \label{tbl-1}}
\tablewidth{0pt}
\tablehead{
\colhead{Parameter}&\colhead{ASCA\tablenotemark{a}}&\colhead{SAX\tablenotemark{a}}&\colhead{SAX\tablenotemark{a}}&\colhead{SAX\tablenotemark{a}}\\&\colhead{1996}&\colhead{1997}&\colhead{1998}&\colhead{2000}\\}
\startdata
N$_{\rm H}$\tablenotemark{b} & 11$^{90}_{-5}$ & 5.7$^{+3.3}_{-0.0}$  & 29$^{+230}_{-23}$ & 6.6$^{+38}_{-0.9}$ \\
Photon index ($\Gamma_1$) & 0.46$^{+0.63}_{-0.37}$ & 0.81 $\pm$ 0.05 & 0.37$^{+0.54}_{-0.38}$ & 0.63 $\pm$ 0.17 \\
Fe line flux\tablenotemark{c}          &  -     & 3.0 $\pm$ 1.2 & - & 2.8 $\pm$ 2.5 \\
log L$_{\rm X}$ \tablenotemark{d}          &  36.09     & 36.78 & 35.48 &  36.18 \\
Reduced $\chi^2$/dof           & 0.23/9           & 1.44/79           & 0.49/20           & 0.55/46  \\

\tablenotetext{a}{GIS2 + GIS3 for ASCA and LECS + MECS2 + MECS3 for SAX }
\tablenotetext{b}{$10^{20}$ atoms cm$^{-2}$}
\tablenotetext{c}{$10^{-5}$ photons cm$^{-2}$ s$^{-1}$}
\tablenotetext{d}{erg s$^{-1}$, 2.0-10.0 keV. Since the absorption
column density is not well constrained, the L$_{\rm X}$ given here is 
not corrected for absorption}
\tablenotetext{e}{$10^{-13}$ erg cm$^{-2}$ s$^{-1}$}

\enddata

\end{deluxetable}

\clearpage

\begin{figure}[ht]
\vskip 7in
\includegraphics{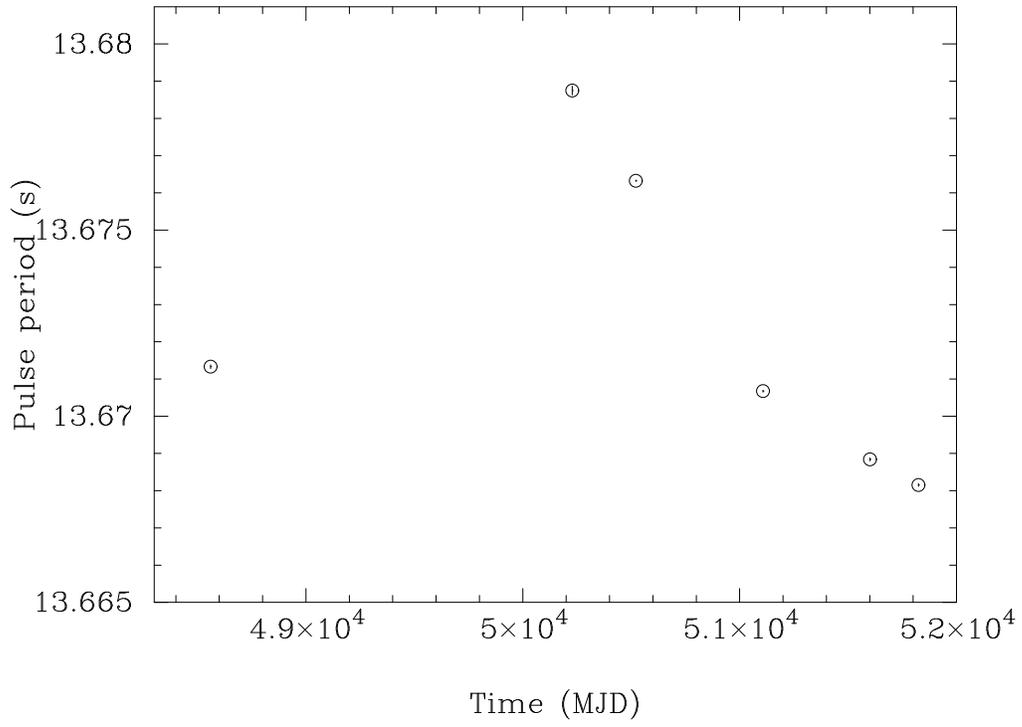}
\caption{Pulse period history of EXO~053109--6609.2. The error bars lie
within the circular symbols.}
\end{figure}

\newpage

\begin{figure}[ht]
\vskip 8in
\includegraphics{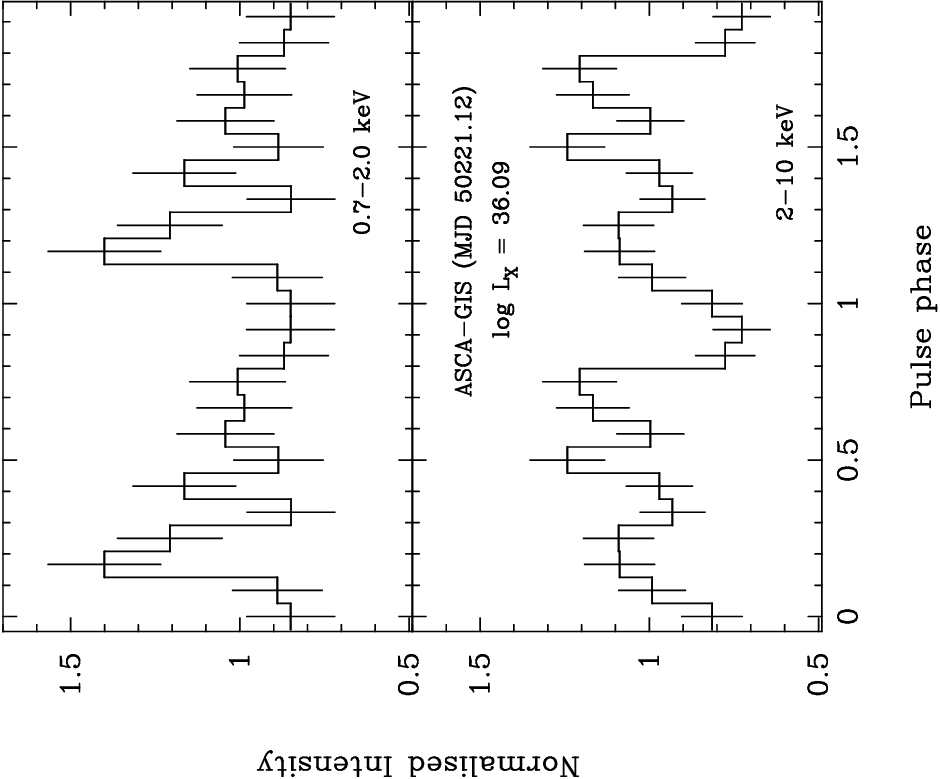}
\includegraphics{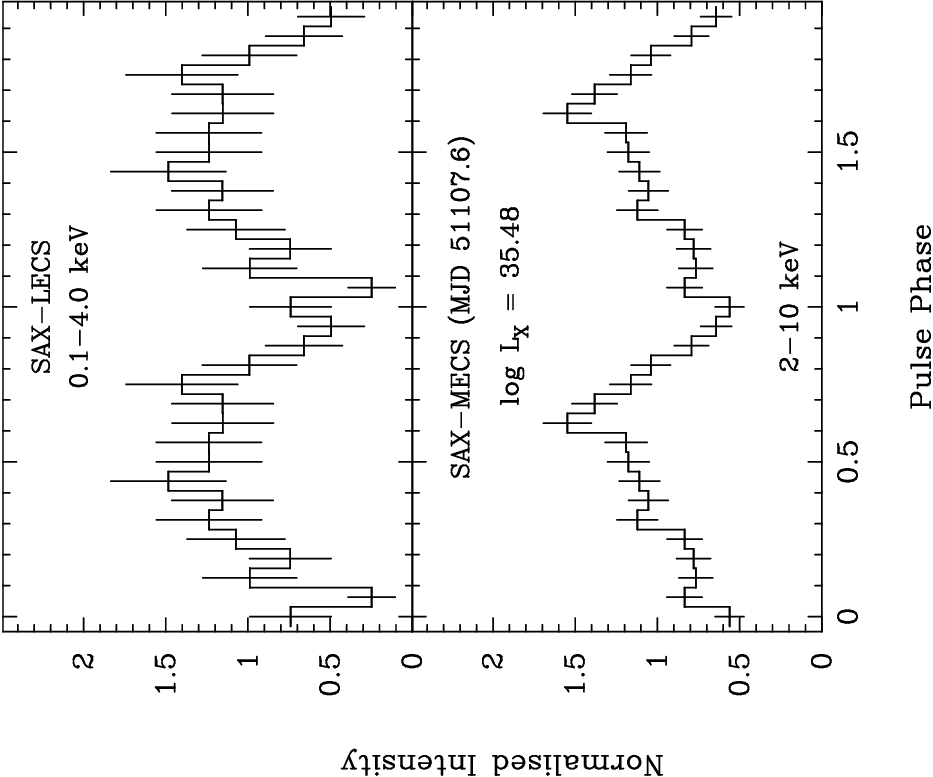}
\includegraphics{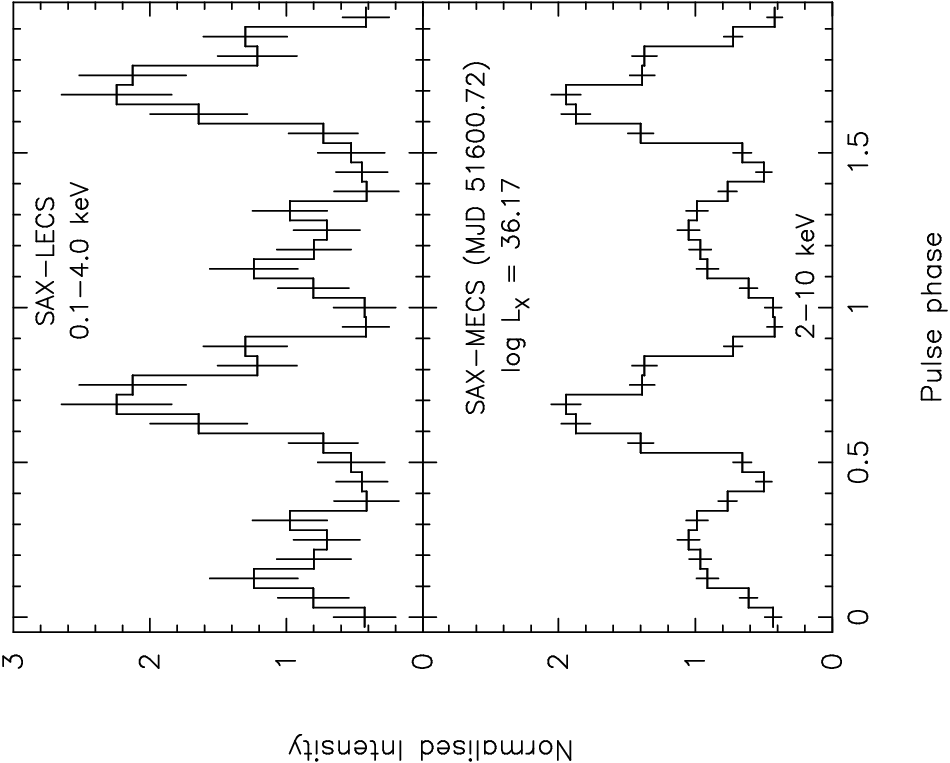}
\caption{Pulse profiles of EXO~053109--6609.2 measured with the ASCA
and Beppo-SAX detectors on various occasions are shown here.}
\end{figure}

\newpage

\begin{figure}[ht]
\vskip 7in
\includegraphics{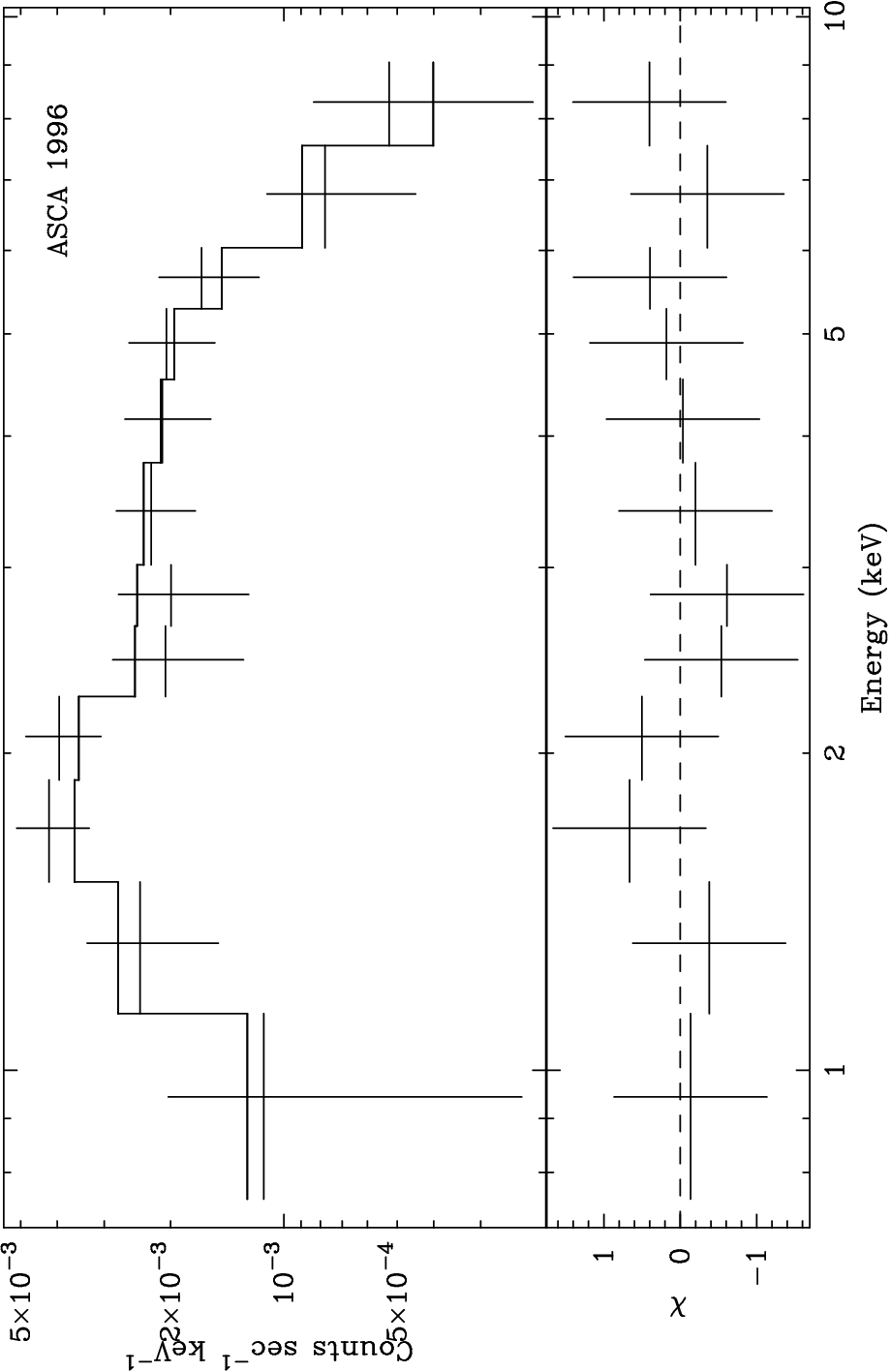}
\includegraphics{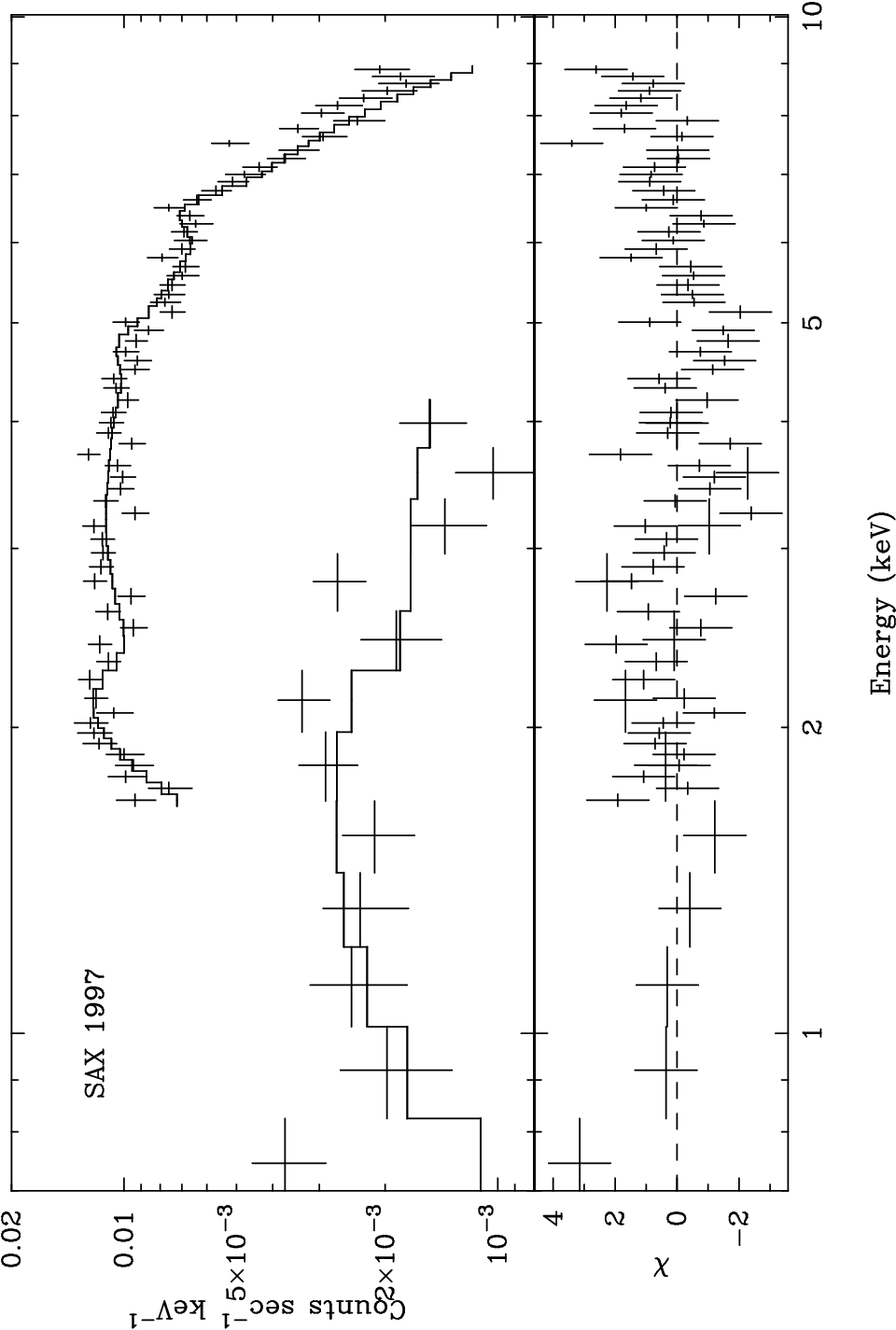}
\includegraphics{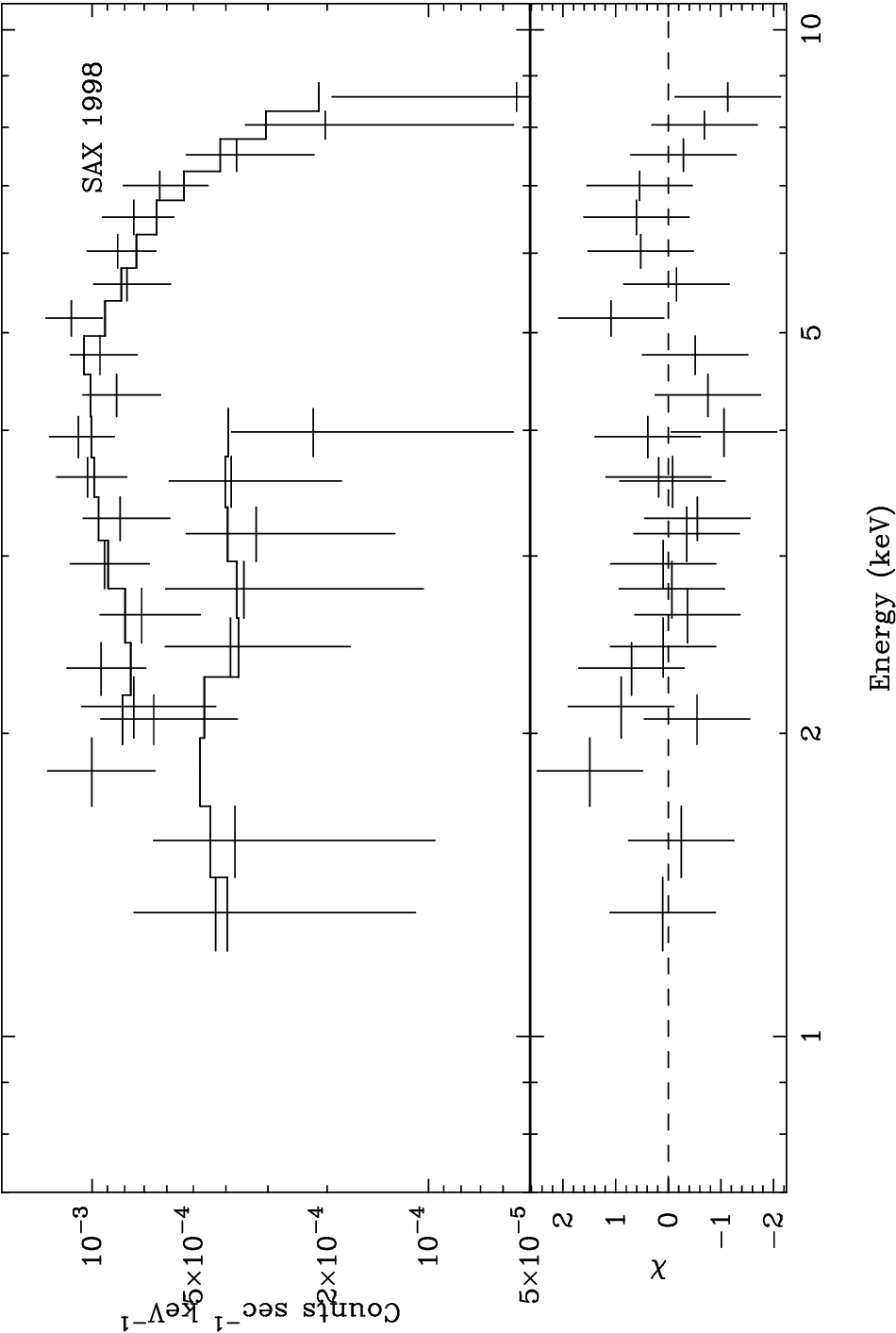}
\includegraphics{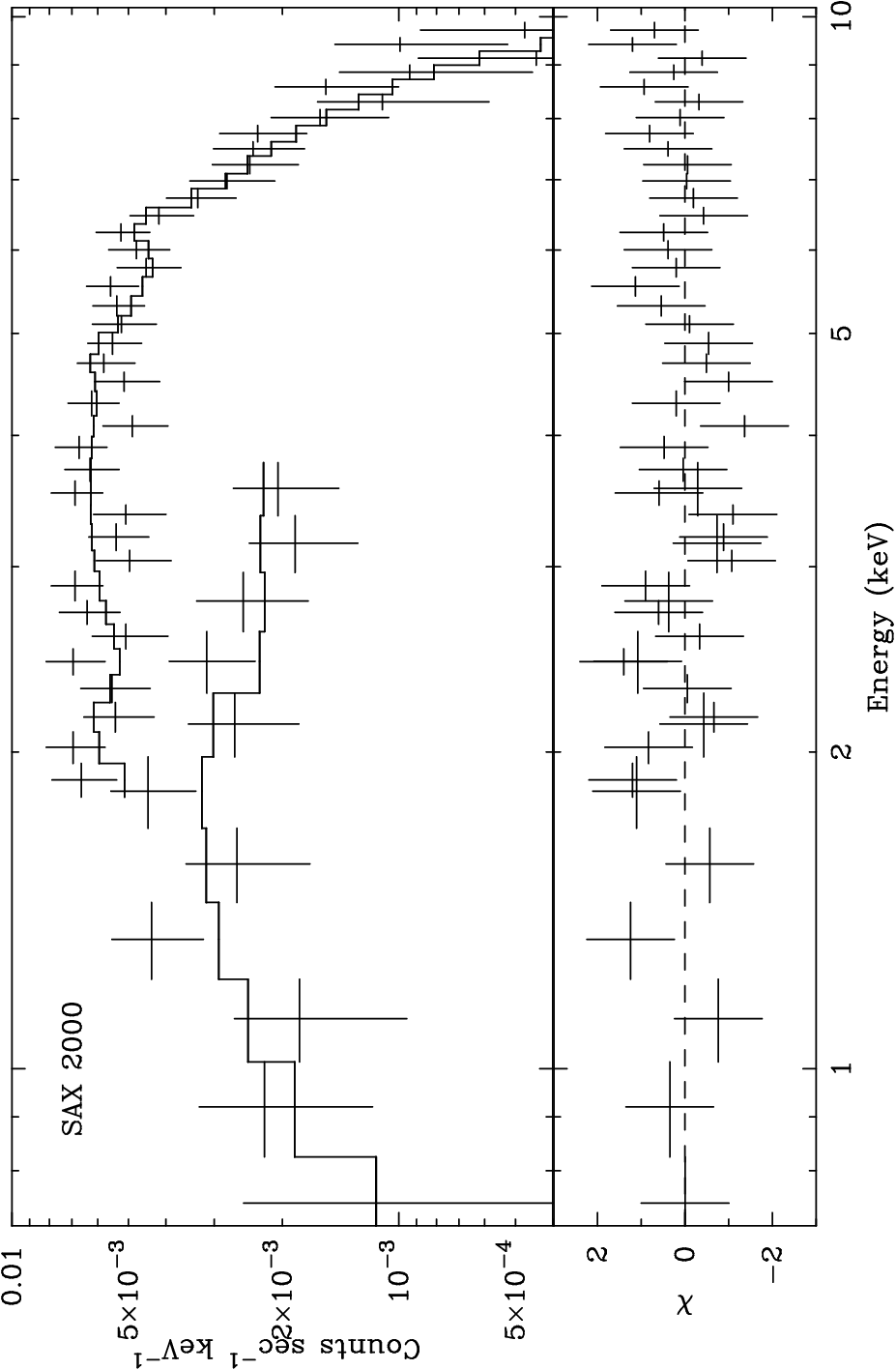}
\caption{Energy spectra of EXO~053109--6609.2 measured with the ASCA and
Beppo-SAX on various occasions along with the best fitted model spectra
and the residuals.}
\end{figure}

\newpage

\end{document}